\documentstyle[11pt,psfig,amssymb] {article}
\input epsf.sty

\newcommand{\be}{\begin{equation}}
\newcommand{\ee}{\end{equation}}
\newcommand{\bd}{\begin{displaymath}}
\newcommand{\ed}{\end{displaymath}}
\newcommand{\ba}{\begin{array}}
\newcommand{\ea}{\end{array}}
\newcommand{\bt}{\begin{tabular}}
\newcommand{\et}{\end{tabular}}
\newcommand{\bc}{\begin{center}}
\newcommand{\ec}{\end{center}}
\newcommand{\bn}{\begin{enumerate}}
\newcommand{\en}{\end{enumerate}}
\newcommand{\bi}{\begin{itemize}}
\newcommand{\ei}{\end{itemize}}
\newcommand{\bqr}{\begin{eqnarray}}
\newcommand{\eqr}{\end{eqnarray}}
\newcommand{\bfig}{\begin{figure}[tbp]}
\newcommand{\efig}{\end{figure}}
\newcommand{\btab}{\begin{table}[ht]}
\newcommand{\etab}{\end{tabular}\ec\end{table}}
\newcommand{\bl}{\begin{large}}
\newcommand{\el}{\end{large}}

\newcommand{\nb}{\nonumber}

\newcommand{\nuc}[2]{\mbox{\relax\ifmmode{}^{#1}{\protect\text{#2}}\else${}^{#1}$#2\fi}}

\title{Goldstone-Brueckner Perturbation Theory \\ 
Extended in Terms of Mixed \\
 Non-Orthogonal Slater-Determinants.}
\author     {
             T. Duguet\footnote{Present address: Argonne National Laboratory, Physics Division, 9700 S. Cass Av., Argonne, IL, 60439, USA. E-mail~: duguet@theory.phy.anl.gov}, \\
             {\em Service de Physique Th\'eorique, CEA Saclay,} \\
             {\em 91191 Gif sur Yvette Cedex, France} \\
            }

\begin{document}

\maketitle

\begin{abstract}

The Goldstone-Brueckner perturbation theory is extended to incorporate in a simple way correlations associated with large amplitude collective motions in nuclei. The new energy expansion making use of non-orthogonal vacua still allows to remove the divergences originating from the hard-core of the bare interaction. This is done through the definition of a new Brueckner matrix summing generalized Brueckner ladders. At the lowest-order, this formalism motivates variational calculations beyond the mean-field such as the Generator Coordinate Method (GCM) and the Projected Mean-Field Method from a perturbative point of view for the first time. Going to higher orders amounts to incorporate diabatic effects in the GCM and to extend the projection technique from product states to well-defined correlated states.



\end{abstract}


\section{Introduction}
\label{intro}

The mean-field approximation relies on a particular choice of the approximate trial-state of the system, namely a Slater-determinant $| \, \Phi^{\alpha}_{0} \rangle$~\cite{blaiz.ripk}~\footnote{In its generalized form taking care of static pairing correlations, the mean-field approximation makes use of a state being a product of independent quasi-particles instead of independent particles.} describing the system as $N$ independent particles. The mean-field approximation can also be viewed as the zero-order approximation of the actual ground-state energy in a perturbative expansion written in terms of the residual interaction.

In the case of nuclear structure, the mean-field approximation fails from the outset because of the strong repulsive core of the bare nucleon-nucleon interaction which leads to divergences when limiting the perturbative expansion to any order in the interaction. However, Brueckner~\cite{brueck} showed that the energy expansion can be reordered as a function of hole-lines number in the graphs, which amounts to an expansion in the density of the system. Such a reordering takes care of two-body short-range correlations induced by the repulsive core of the nucleon-nucleon interaction. It leads to an expression of the energy in terms of a renormalized interaction $G$. Using it, one can study the lowest-order approximation and define a meaningful mean-field picture. One can then evaluate the improvements achieved by including higher order diagrams since each of them is well-behaved.

In many-body systems, several ways to go beyond the mean-field approximation exist, depending upon the physical situation of interest~\cite{blaiz.ripk}. Without particularly relating it to any perturbative expansion, one can improve the approximate trial wave-function of the system in a variational picture in order to include correlations in the ground-state and calculate excited states. The projected mean-field method~\cite{yocco,diet} and the Generator Coordinate Method (GCM)~\cite{hill,bonche1,bonche2} are such examples. The correlations considered in these two methods are those associated with large amplitude collective motions which are known to be very important in low-lying nuclear structure. This is done by using a trial state written as a superposition of several non-orthogonal product functions referring to different mean-fields:

\begin{equation}
| \, \Psi_{k} \rangle \, = \, \sum_{\alpha} f^{k}_{\alpha} \, | \, \Phi^{\alpha}_{0} \rangle \, \, \, \, \, \, .  \label{mixing}
\end{equation}

Once such a trial state is given, its mean energy:

\begin{eqnarray}
{\cal E}_{k}^{mix}  &\equiv&  \, \, \, \, \, \, \, \,  \, \, \frac{\langle \, \Psi_{k} \, | \, H \, | \, \Psi_{k} \rangle}{\langle \, \Psi_{k} \, | \, \Psi_{k} \rangle} \, \, \, \,  \, \, ,  \label{factorisee} \\
\nb \\
&=& \frac{\sum_{\alpha,\beta} \, f_{\beta}^{k \, \ast} f^{k}_{\alpha} \langle \, \Phi^{\beta}_{0} \, | \, H \, | \, \Phi^{\alpha}_{0}  \rangle}{\sum_{\alpha,\beta} \, f_{\beta}^{k \, \ast} f^{k}_{\alpha} \langle \, \Phi^{\beta}_{0} \, | \, \Phi^{\alpha}_{0}  \rangle} \, \, \, \, \, \, .  
\label{energy1}
\end{eqnarray}
can be minimized with respect to variational parameters. The mean-field approximation is recovered if a single coefficient $f^{k}_{\alpha}$ is non zero; its self-consistent version corresponding to the minimization of the energy with respect to individual wave-functions. 

A problem is that going beyond the mean-field through Eq.~\ref{mixing}-\ref{energy1}, it is no longer possible to connect the variational calculation to any existing diagrammatic picture. Indeed, pertubation theories developed up to now can only deal with small amplitude correlations, these correlations being short-ranged or long-ranged~\cite{blaiz.ripk}. The treatment of large amplitude collective motions as defined by Eq.~\ref{mixing}-\ref{energy1} makes use of non-orthogonal vacua connected through specific infinite sums of particle-hole excitations which can hardly be identified in usual expansions. To make such a link is a question of interest since the systematic character of a perturbative expansion is a very powerful tool in order to classify the correlations included or forgotten in a given variational calculation. An appropriate perturbative scheme would also be useful to identify the effective interaction which aims at renormalizing two-body correlations in the context of large amplitude collective motions.

The aim of the present paper is to extend the Brueckner-Goldstone theory in order to motivate the methods embodied by Eq.~\ref{mixing}-\ref{energy1} from a perturbative point of view. The lowest-order approximation of the new expansion provides an energy having the form as given by Eq.~\ref{energy1} and thus describes large amplitude collective motions\footnote{We concentrate on perturbation theories written in a Hilbert space with definite particle number. In other words, only normal contractions $\langle c^{\dagger} \, c \rangle$ are non-zero. Such theories aim at treating the particle-hole channel of the interaction.}. Higher-orders introduce more correlations and of different nature than in the lowest-order. Namely, they correspond to the introduction of non-adiabatic effects in the coupling of individual and collective degrees of freedom. Moreover, the perturbative scheme allows to eliminate divergences arising from the strong repulsive core of the interaction by summing Brueckner ladders in the context of mixed vacua. It thus provides a microscopically defined effective interaction to be used in Eq.~\ref{energy1}.

The present work is organized as follows: we first remind some important features of the usual Goldstone-Brueckner perturbation theory in section~\ref{secgoldstone}. Then we develop the new perturbative scheme in section~\ref{generalization} where we define and study a generalized Brueckner $G$ matrix. In sections~\ref{gcm} and ~\ref{proj}, the link between the lowest-order approximation and the configurations mixings used in variational calculations is illustrated for the GCM and the projected mean-field method. The conclusions are given in section~\ref{conclu}.

\section{Standard Goldstone-Brueckner Theory.}
\label{secgoldstone}

\subsection{Ground-State Expansion.}
\label{goldformula}

Let us start with the Hamiltonian of the $N$ interacting nucleons. It consists of a kinetic energy term and a two-body force $V$, which includes a realistic nucleon-nucleon interaction and the Coulomb interaction:
\begin{equation}
H \, = \,  \, \, t  \, \, +  \, \, \, V \, \, \, \, . 
\label{H}
\end{equation}

We define a one-body Hamiltonian $h_{0}^{\alpha}$ potentially breaking symmetries of $H$. Its ground-state and excited states are denoted $| \, \Phi^{\alpha}_{0} \rangle$ and $| \, \Phi^{\alpha}_{i} \rangle \,$ respectively. We make the additional hypothesis that $| \, \Phi^{\alpha}_{0} \rangle$ is non-degenerate. One specifies this mean-field through:

\begin{equation}
h_{0}^{\alpha} \, \, = \, \, t  \, +  \, \Gamma^{\alpha} \, \, = \, \, \sum_{n} \, \epsilon_{\alpha_{n}} \, \alpha^{\dagger}_n \, \alpha_n \, \,  \, \, \, , \label{h}
\end{equation}

\begin{equation} 
h_{0}^{\alpha} \, \, \, | \, \Phi^{\alpha}_{i} \rangle \, \, =  \, \, {\cal E}_{i}^{\alpha} \,  \, | \, \Phi^{\alpha}_{i} \rangle \, \, \, \, \, ,  \label{h'}
\end{equation}
where the $\left\{\alpha\right\}$ single-particle basis is chosen as diagonalizing $h_{0}^{\alpha}$. For our purpose, we assume the index $i$ to be discrete and the ${\cal E}_{i}^{\alpha}$ to be ordered with increasing value. The exact Hamiltonian can be related to $h_{0}^{\alpha}$ through:

\begin{eqnarray}
H \, &=& \, h_{0}^{\alpha}  \, \, +  \, \, \, h_{1}^{\alpha} \, \,  \, \, \, , \nb
\end{eqnarray}
where
\begin{equation}
h_{1}^{\alpha} \, = \,  V   \, \, -  \, \, \, \Gamma^{\alpha}  \, \,  \, \, \, ,
\label{h1}
\end{equation}
is characterized as the residual interaction.

In what follows, we make use of the Gell-Mann-Low adiabatic theorem~\cite{gell} which states that an eigenstate of $H$ is obtained when evolving $| \, \Phi^{\alpha}_{0} \rangle$ adiabatically from $t = -\infty$ to $t = 0$. Actually, no rule exists saying that this eigenstate is necessarily the ground-state of the full Hamiltonian. We will consider that it is so~\cite{fet}. This theorem explicitly reads as:

\begin{eqnarray}
| \, \Theta^{\alpha}_{0}  \rangle \, &=& \lim_{\epsilon \, \rightarrow \, 0} \frac{U^{\alpha}_{\epsilon} (0,-\infty) \, | \, \Phi^{\alpha}_{0} \rangle}{\langle \, \Phi^{\alpha}_{0}\, | U^{\alpha}_{\epsilon} (0,-\infty) | \, \Phi^{\alpha}_{0} \rangle }  \, \, \, \, \, \, \, . \label{gell}
\end{eqnarray}

In $| \, \Theta^{\alpha}_{0}  \rangle \,$, the label $\alpha$ reminds from which unperturbed state $| \, \Phi^{\alpha}_{0} \rangle$ the actual ground-state comes from. The adiabatic evolution operator $U^{\alpha}_{\epsilon}$ in the $\alpha$ interaction representation~\cite{fet} is:

\begin{eqnarray}
U^{\alpha}_{\epsilon} (t,t_0) \, &=& \,   e^{i h^{\alpha}_{0} t  / \hbar} \, \, U_{\epsilon} (t,t_0)  \, \,  e^{- i h^{\alpha}_{0} t_0  / \hbar}  \nb \\
&& \label{Uevol}  \\
&=& \, e^{i h^{\alpha}_{0} t  / \hbar} \, \, e^{- i \int_{t_0}^{t}H(\epsilon,\tau) d\tau  / \hbar}  \, \,  e^{- i h^{\alpha}_{0} t_0  / \hbar} \nb \, \, \, \, \, ,
\end{eqnarray}
where $U_{\epsilon}$ is the evolution operator in the Schr{\oe}dinger representation. The derivation of Eq.~\ref{gell} relies on two fundamental properties of quantum mechanics~\cite{fet,messiah}:

\bi
\item[(1)] Making an eigenstate of $H (t_0)$ to evolve adiabatically from $t = t_0$ through the evolution operator of the system drives the system to an eigenstate of $H (t_1)$ at $t = t_1$. \\
\item[(2)] It is possible to find an adiabatic time-dependent Hamiltonian $H (\epsilon,t)$ which reduces to the independent particle Hamiltonian $h^{\alpha}_0$ at $t_0 = -\infty$ and which equals the full Hamiltonian $H$ of the system at $t_1 = 0$. 
\ei

In the present case, the auxiliary time-dependent Hamiltonian switching the perturbation $h^{\alpha}_{1}$ adiabatically is given by:

\begin{equation}
H (\epsilon,t) \, = \, h^{\alpha}_{0} \, +  \, e^{- \epsilon |t|} \, \, h^{\alpha}_{1} \, \, \, \, \, \, ,
\label{hepsilon}
\end{equation}
where $\epsilon$ is a small positive parameter. At very large times, both in the past and in the future, this Hamiltonian effectively reduces to $h^{\alpha}_{0}$ which presents a known solution, whereas at $t = 0$, $H (\epsilon,t)$ becomes the full Hamiltonian of the interacting system.  The limit $\epsilon \, \rightarrow \, 0$ is taken at the end and the results are independent of $\epsilon$. Note that this method does not rely on the application of the evolution operator corresponding to the actual Hamiltonian $H$ on a given state ``at $t_0 = -\infty$''. 

The normalization of the actual ground-state $| \, \Theta^{\alpha}_{0}  \rangle \,$ is $\langle \, \Phi^{\alpha}_{0}\, | \, \Theta^{\alpha}_{0}  \rangle \, = \, 1$. It is chosen through the denominator of Eq.~\ref{gell} which exactly cancels disconnected vacuum/vacuum diagrams in the numerator~\cite{gold}. This is crucial as both the numerator and the denominator diverge in the limit $\epsilon \, \rightarrow \, 0$.

Performing explicitly the time integrations contained in $U^{\alpha}_{\epsilon}$, Goldstone~\cite{gold} found a linked expansion of $| \, \Theta^{\alpha}_{0}  \rangle $ allowing for a diagrammatic representation and reading as:

\begin{equation}
| \, \Theta^{\alpha}_{0}  \rangle \, = \, \sum_{n}^{} \, \left(\frac{1}{{\cal E}^{\alpha}_{0} \, - \, h^{\alpha}_{0}} \, h^{\alpha}_{1} \right)^n \, | \, \Phi^{\alpha}_{0} \rangle_{Linked} \, \, \, \, \, \, \, ,
\label{goldstone}
\end{equation}
where $Linked$ means that $| \, \Phi^{\alpha}_{0} \rangle$ does not occur as an intermediate state in the diagrams.

\subsection{Brueckner Ladders.}
\label{bruechladders}

\begin{figure}
\begin{center}
\leavevmode
\centerline{\psfig{figure=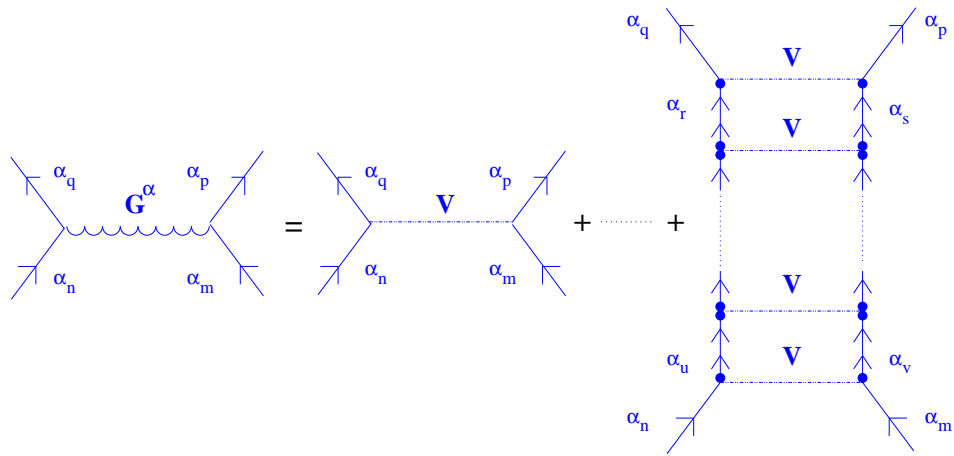,height=6cm}}
\end{center}
\caption{$G^{\alpha}$ matrix elements summing p-p ladders with respect to the vacuum $| \, \Phi^{\alpha}_{0} \rangle$. In any ladder, all $V$ interactions (except possibly the first and last) are drawn between two up-going lines simulating particle states outside the Fermi sea. The rest of the diagram is not allowed to have any interaction between the first and the last interaction in the ladder. Points appearing in the present figure are superfluous here, but will become essential in the following. They will be defined in Fig.~\ref{defgraph}.}
\label{diaggmatrix}
\end{figure}

Following Brueckner~\cite{brueck}, each $V$ involved in a linked diagram of Eq.~\ref{goldstone} is to be replaced by a reaction matrix $G^{\alpha}$ summing particle-particle (p-p) ladders in the $\left\{ \alpha \right\}$ single-particle basis. This removes the strong repulsive core of the bare nucleon-nucleon interaction which makes the expansion in terms of $V$ irrelevant. The replacement of $V$ by $G^{\alpha}$ is possible because each time a $V$ interaction occurs in the original expansion of $| \, \Theta^{\alpha}_{0}  \rangle$, one can find the infinite set of identical diagrams except that a succession of $V$ interactions connecting particle states occurs instead of the original $V$ and this, before any interaction takes place somewhere else in the graph. The corresponding diagrammatic content is shown on Fig.~\ref{diaggmatrix}. This matrix satisfies a self-consistent equation of the form:

\begin{eqnarray}
G^{\alpha} (W_{\alpha}) \, &=& \, V  \, \, \, + \, \, \, V \, \, \, \frac{Q^{\alpha}}{W_{\alpha} \, \, - \, \, h^{\alpha}_{0}} \, \, G^{\alpha} (W_{\alpha})  \, \, \, \,  \, \,  ,  \label{gmatrix} 
\end{eqnarray}
with~\footnote{The two-body states $|\, \alpha_{p} \alpha_{p'}  \rangle$ are non-antisymmetrized here.}
\begin{equation}
Q^{\alpha} \, = \, \sum_{\epsilon_{\alpha_{p}},\epsilon_{\alpha_{p'}}  > \epsilon_{F}^{\alpha}}^{} \, |\, \alpha_{p} \alpha_{p'}  \rangle \,  \langle \, \alpha_{p} \alpha_{p'} \, |  \, \, \, \, \, \, \, \,  \, \,  .
\label{pauli}
\end{equation}

The Pauli operator $Q^{\alpha}$ acts in the two-particle space by excluding occupied states in $| \, \Phi^{\alpha}_{0} \rangle \,$ as intermediate states. The $W_{\alpha}$ dependence of $G^{\alpha}$ denotes that the in-medium interaction of two particles depends on the energy of the others during the interaction. The $G^{\alpha} (W_{\alpha})$ matrix elements together with the precise definition of $W_{\alpha}$ are given in appendix A.1. Iterating Eq.~\ref{gmatrix} shows that $G^{\alpha}$ takes into account the interaction between two particles to all orders of the potential in presence of the other nucleons. 

Let us now replace each $h^{\alpha}_{1}$ by $G^{\alpha}_{1} (W_{\alpha}) = G^{\alpha} (W_{\alpha}) -  \Gamma^{\alpha}$ in Eq.~\ref{goldstone}. All diagrams with two successive $G^{\alpha}$ interactions joined through particle states are excluded from the sum in order to avoid double counting; the sum runs over linked irreducible diagrams~\cite{gold}: 

\begin{eqnarray}
| \, \Theta^{\alpha}_{0}  \rangle \, &=& \,  \sum_{n}^{} \, \left(\frac{1}{{\cal E}^{\alpha}_{0} \, - \, h^{\alpha}_{0}} \, G^{\alpha}_{1} (W_{\alpha}) \right)^n \, | \, \Phi^{\alpha}_{0} \rangle_{Linked/Irred.} \, \, \, \, \, \, \, .  \label{extended} 
\end{eqnarray}

All graphs are generated in this way. This is nothing but a reordering of diagrams. Each term of the perturbative expansion is now finite and ``well-behaved'' even if the original interaction contains a strong short-range repulsion as in the nuclear case~\cite{ring1}. The Brueckner scheme has to be seen as the minimum recipe to get rid of the hard core problem. More elaborate treatments of the many-body problem through systematic renormalization of propagators and vertices exist, using for instance the self-consistent Green function approach~\cite{dick,mut}.

\subsection{One-body Potential $\Gamma^{\alpha}$.}
\label{onebodypot}

We do not discuss in detail the question of the most appropriate one-body potentials $\Gamma^{\alpha}$ to be chosen for p-p, p-h and h-h matrix elements~\cite{mut,kohl}. Let us just recall few points. First, one can define a fixed, purely phenomenological one-body potential $\Gamma^{\alpha}$. Such a choice suffers from its non-systematic nature. Another choice is to define $\Gamma^{\alpha}$ in analogy to the Hartree-Fock potential with the bare interaction $V$ replaced by the $G^{\alpha}$ matrix. This is the Brueckner-Hartree-Fock (BHF) approximation. It reads:

\begin{figure}
\begin{center}
\leavevmode
\centerline{\psfig{figure=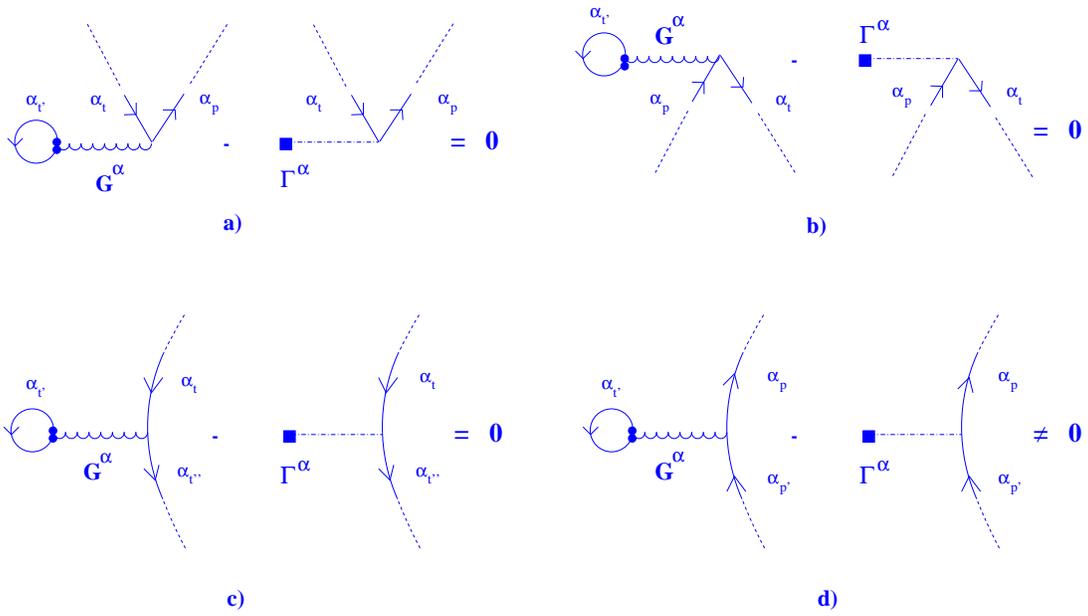,height=8cm}}
\end{center}
\caption{Cancelled diagrams in $| \, \Theta^{\alpha}_{0}  \rangle$ in relation with the Brueckner-Hartree-Fock definition of the one-body potential $\Gamma^{\alpha}$. Dashed lines mean that the remaining part of the summed graphs are identical.}
\label{defU}
\end{figure}

\begin{equation}
\langle  \, \alpha_{i} \, | \, \Gamma^{\alpha} \, | \, \alpha_{j} \,  \rangle \, =  \, \sum_{\alpha_{k}, \alpha_{k'}} \, \langle  \, \alpha_{i} \, \alpha_{k} \, | G^{\alpha} (0) | \, \alpha_{j} \, \alpha_{k'} \,  \rangle \, \rho^{\alpha}_{\alpha_{k} \alpha_{k'}} \, \, \, \, \, , 
\label{onebod}
\end{equation}
for h-h, p-h and p-p matrix elements, where the one-body density matrix $\rho^{\alpha}$ associated with $| \, \Phi^{\alpha}_{0} \rangle$ is defined in any single-particle basis $\left\{c_{i}\right\}$ by:

\begin{equation}
\rho^{\alpha}_{c_{s}c_{r}} \, = \, \langle  \, c_{s} \, | \, \hat{\rho}^{\alpha} | \, c_{r} \,  \rangle \, = \,\frac{\langle  \, \Phi^{\alpha}_{0} | \, c^{\dagger}_r \, c_s | \, \Phi^{\alpha}_{0} \rangle}{\langle  \, \Phi^{\alpha}_{0} \, | \, \Phi^{\alpha}_{0} \rangle} \, \,  \, \, \, \, \, ,
\label{densdiago}
\end{equation}
and is diagonal in the $\{\alpha_{k}\}$ basis in such a way that $\rho^{\alpha}_{\alpha_{k} \alpha_{k'}} = \delta_{kh} \, \delta_{k'h}$, where $h$ denotes hole states.

In Eq.~\ref{onebod}, $G^{\alpha}$ is taken {\it on the energy shell} ($W_{\alpha} = 0$). Such a definition allows to cancel a large number of diagrams in the wave-function. This is schematically shown on Fig.~\ref{defU}. Diagrams~\ref{defU}a and ~\ref{defU}b correspond to insertions on vertices. They involve a $G^{\alpha}$ interaction with passive unexcited states and a $\Gamma^{\alpha}$ p-h interaction. Diagrams~\ref{defU}c embody insertions on hole lines. These three groups of diagrams have been shown to cancel thanks to the generalized time ordering~\cite{thoul,bethe} although some of the $G^{\alpha}$ interactions are originally not on the energy shell ($W_{\alpha} \neq 0$). On the contrary, insertions on particle lines cannot be systematically cancelled by any definitions of $\langle \alpha_{p} \, | \, \Gamma^{\alpha} \, | \, \alpha_{p'}  \rangle$ as depicted on Fig.~\ref{defU}d. The two parts of these diagrams have to be calculated explicitely when appearing. The choice of the p-p matrix elements of $\Gamma^{\alpha}$ has been of considerable controversy but we will not go into more details here and refer to Ref.~\cite{becker,mac,trip,sau,song}. 

Finally, let us say that the BHF choice for $\Gamma^{\alpha}$ is not the most appropriate one from the quantitative point of view~\cite{kohl,davies1,nege3}. A definition including higher-order diagrams is known to take care of the rearrangement potential~\cite{davies2} which significantly improves density distributions in heavy nuclei as well as the agreement between theoretical single-particle and separation energies~\cite{nege3,becker2}. Indeed, higher-order graphs are cancelled with such a choice in a way that the zero-order is a better approximation for observables such as the one-body density matrix determining one-body observables. 

\subsection{Energy Expansion.}
\label{enerexpansion1}

The Gell-Mann-Low expression for the energy $E_{0} = {\cal E}_{0}^{\alpha} + \Delta {\cal E}_{0}^{\alpha}$ of the interacting ground-state $| \, \Theta^{\alpha}_{0}  \rangle$ is obtained through the scalar product of $H \, | \, \Theta^{\alpha}_{0}  \rangle$ with the unperturbed bra $\langle \, \Phi^{\alpha}_{0}|$ and reads as:

\begin{equation}
\Delta {\cal E}_{0}^{\alpha} \, = \, \lim_{\epsilon \, \rightarrow \, 0} \frac{\langle \, \Phi^{\alpha}_{0}| \, h^{\alpha}_{1} \, U^{\alpha}_{\epsilon} (0,-\infty) \, | \, \Phi^{\alpha}_{0} \rangle}{\langle \, \Phi^{\alpha}_{0}\, | U^{\alpha}_{\epsilon} (0,-\infty) | \, \Phi^{\alpha}_{0} \rangle } \, \, \, \, \, \, .
\label{gellenergy}
\end{equation}

The Goldstone-Brueckner linked/irreducible expansion deduced from it is:

\begin{equation}
\Delta {\cal E}_{0}^{\alpha} \, = \, \sum_{n}^{} \, \langle \, \Phi^{\alpha}_{0}| \, G^{\alpha}_{1} (W_{\alpha}) \, \left(\frac{1}{{\cal E}^{\alpha}_{0} \, - \, h^{\alpha}_{0}} \, G^{\alpha}_{1} (W_{\alpha}) \right)^n \, | \, \Phi^{\alpha}_{0} \rangle_{Linked/Irred.} \, \, \, \, \, \, ,
\label{goldenergy}
\end{equation}
where the sum runs now over all connected graphs with no external lines. Finally, a meaningful mean-field approximation in nuclear-structure is deduced by considering the lowest-order in terms of $G^{\alpha}$ in Eq.~\ref{goldenergy}. This is the Brueckner-Hartree-Fock approximation:

\begin{eqnarray}
E^{\, n=0}_{0} \, &=& \, {\cal E}_{0}^{\alpha} \, + \, \langle \, \Phi^{\alpha}_{0} | \, \, G^{\alpha}_{1} (0) \, \, | \,  \Phi^{\alpha}_{0}  \rangle \nb \\
&& \nb \\
&=& \langle \, \Phi^{\alpha}_{0} | \, \, t \, + \, G^{\alpha} (0) \, \, | \,  \Phi^{\alpha}_{0}  \rangle \label{middleenergy} \\
&& \nb \\
&=& \, \sum_{\alpha_{h}} \, t_{\alpha_{h}\alpha_{h}} \,  \rho^{\alpha}_{\alpha_{h}\alpha_{h}}  \,+ \, \frac{1}{2} \sum_{\alpha_{h}\alpha_{h'}} \, \bar{G}^{\alpha}_{\alpha_{h}\alpha_{h'}\alpha_{h}\alpha_{h'}} (0) \, \rho^{\alpha}_{\alpha_{h}\alpha_{h}} \,  \rho^{\alpha}_{\alpha_{h'}\alpha_{h'}} \, \, \, \, \, ,
\nb
\end{eqnarray}
where $\bar{G}^{\alpha}_{\alpha_{m}\alpha_{p}\alpha_{n}\alpha_{q}} (0)$ are antisymmetrized matrix elements taken on the energy shell. One can effectively see that Eq.~\ref{middleenergy} formally has the form of a mean-field energy for the effective ``Hamiltonian'' $t \, + \, G^{\alpha} (0)$.

We will not go into more details concerning the standard Brueckner theory and properties of the Brueckner $G$ matrix. For extensive presentations, we refer to Ref.~\cite{brueck,gold,kohl,day}. Let us just mention that the renormalization of short-range correlations associated with the exchange of $\omega$ vector mesons is not the only significant origin of in-medium effects and corresponding density-dependence. For instance, the exchange of pions should be important. In fact, the Brueckner summation originally introduced as the minimal recipe to remove the hard-core problem provides in-medium effects, not only from the short-range repulsive part, but also from the long-range part, i.e from the tensor part of the bare force~\cite{kohl1,kohl2}. This is due to the dependence of the tensor force contribution to the energy on the Pauli operator $Q^{\alpha}$. This suggests that a part of the in-medium effects generated by pions is taken into account through the ladder summation (the ladders cover the whole energy range from low to high excitations energies). Evidently, such a perturbation theory based on nucleon-nucleon potentials (possibly including three-body ones) neglects the dynamical mesonic degrees of freedom. One should go back to richer treatments of the many-body problem including these degrees of freedom explicitely in order to evaluate quantitatively their full in-medium effects~\cite{weise}.

\section{Extended Perturbation Theory.}
\label{generalization}

The previous section reminded the Brueckner-Goldstone expansion referring to a given mean-field unperturbed state $| \, \Phi^{\alpha}_{0} \rangle$. As said in the introduction, variational calculations relying on the mixing of non-orthogonal product states cannot be motivated by such a perturbative scheme. We thus develop a perturbation theory based on a superposition of non-orthogonal Slater-determinants. A natural idea would be to evolve a linear combination as given by Eq.~\ref{mixing} from  $t = -\infty$ to $t = 0$. To do so, one has to find a one-body Hamiltonian $h^{mix}_0$, whose ground-state is the mixing of non-orthogonal Slater-determinants, and define an auxiliary Hamiltonian of the form:

\begin{equation}
H (\epsilon,t) \, = \, h^{mix}_{0} \, + \, e^{- \epsilon |t|} \, \, \left(H  - h^{mix}_{0} \right) \, \, \, \, \, .
\label{hepsilon2}
\end{equation}

However, it is not possible to find $h^{mix}_0$ for a general configuration mixing such as the one defined by Eq.~\ref{mixing}. To avoid this difficulty, we use a different approach based on the arbitrariness of the unperturbed Hamiltonian $h^{\alpha}_{0}$ with respect to which the ground-state of the system is developed in the standard Goldstone picture. This arbitrariness, which is usually used to optimize the unperturbed state through the ``best'' choice of $\Gamma^{\alpha}$, is used here to superpose the expressions of the actual ground-state obtained starting from several unperturbed Slater-determinants $| \, \Phi^{\alpha}_{0} \rangle$.

\subsection{Ground-State Expansion.}
\label{newgoldformula}

Let us choose {\it a set} of mean-fields, corresponding residual interactions, associated non-degenerate ground-state and excited states $\left\{h^{\alpha}_{0}, h^{\alpha}_{1}, | \, \Phi^{\alpha}_{0} \rangle, | \, \Phi^{\alpha}_{i} \rangle\right\}$. For instance, $\{h_{0}^{\alpha}\}$ may correspond to a set of identical deformed mean-fields with different intrinsic orientations. Their deformation corresponding to the minimum of the associated BHF energy, each of them has a minimal density of individual states around the Fermi energy. This makes any particle-hole (p-h) excitation to cost some energy and each ground-state $| \, \Phi^{\alpha}_{0} \rangle$ to be non-degenerate. A second case of interest consists in defining  $\{h_{0}^{\alpha}\}$ through constrained mean-field calculations along a collective deformation path. In this case, single-particle level crossings may occur at the Fermi level for some values of the constrained deformation $\alpha$. This makes the corresponding $| \, \Phi^{\alpha}_{0} \rangle$ degenerate. Thus, one has to pick up a discrete number of $\alpha$ values along the deformation path for which this does not occur.

The rationale of the present derivation is to make each of these vacua evolve independently from $t = -\infty$ to $t = 0$ as processed in section~\ref{secgoldstone}. The normalization $\langle \, \Phi^{\alpha}_{0} \, | \, \Theta^{\alpha}_{0}  \rangle \, = \, 1$ eliminating disconnected vacuum/vacuum diagrams is chosen for each $\alpha$. In this way, the $| \, \Theta^{\alpha}_{0}  \rangle$ simply differ through their norm. For each $\alpha$, the standard Brueckner scheme is performed in order to get the ground-state wave-function under the form given by Eq.~\ref{extended}. Finally, we perform an arbitrary linear combination of the states $| \, \Theta^{\alpha}_{0}  \rangle$:

\begin{equation}
| \, \Theta_{0} \rangle \, =  \, \sum_{\alpha}^{} \, f_{\alpha} \, | \, \Theta^{\alpha}_{0}  \rangle \, = \, \sum_{\alpha} \, f_{\alpha} \, \sum_{n_{\alpha}}^{} \, \left(\frac{1}{{\cal E}^{\alpha}_{0} \, - \, h^{\alpha}_{0}} \, G^{\alpha}_{1} \right)^{n_{\alpha}} \, | \, \Phi^{\alpha}_{0} \rangle_{Linked/Irred.} \, \, \, \, \, \, \, .
\label{extended2}
\end{equation}

As each $| \, \Theta^{\alpha}_{0}  \rangle \,$ is the ground-state of $H$ with the eigenvalue $E_{0}$, $| \, \Theta_{0}  \rangle \,$ is also the eigenstate with the same eigenenergy. This is true whatever the coefficients $f_{\alpha}$ are. Actually, only the normalization changes from Eq.~\ref{goldstone} to Eq.~\ref{extended2}. This last equation constitutes the starting point of our extended perturbative scheme. The linear superposition in Eq.~\ref{extended2} seems to be redondant and trivial but it provides richer approximate states at any order of the expansion. In particular, the lowest order provides with a superposition of non-orthogonal product states as a well defined approximation of the actual ground-state of the interacting system. In the following, we will focus on the perturbative expansion of the corresponding eigenenergy.

\subsection{Energy Expansion.}
\label{expansion}

In the standard formulation, the starting point the energy expansion is Eq.~\ref{gellenergy} which gives the diagrammatic development of the correction $\Delta {\cal E}_{0}^{\alpha}$ to the unperturbed energy ${\cal E}_{0}^{\alpha}$. It is obtained through the scalar product of $h^{\alpha}_{1} \, | \, \Theta^{\alpha}_{0}  \rangle$ with the unperturbed bra $\langle  \Phi^{\alpha}_{0}  |$. In our extended scheme however, there is no well defined unperturbed state and unperturbed energy since each of the vacuum $| \, \Phi^{\alpha}_{0} \rangle$ evolves independently from $t = -\infty$ before the linear combination is performed. To avoid this ambiguity, we keep the symmetry between the bra and the ket and directly express the total energy $E_{0}$ through the mean value of $H$ in the ground-state $| \, \Theta_{0}  \rangle$. Using Eq.~\ref{extended2}, we get:

\begin{eqnarray}
&& \langle \, \Theta_{0}|  \, H \, | \, \Theta_{0} \rangle \, \, \,  =  \, \, \, E_{0} \, \langle \, \Theta_{0}\, | \, \Theta_{0} \rangle \, \, \, \, \, \, \, \, \,  \nb \\
&& \nb \\
&& \hspace{2.6cm} = \,  \sum_{\alpha,\beta,n_{\alpha},n_{\beta}}  f_{\beta}^{\ast}  f_{\alpha} \, \langle \Phi^{\beta}_{0} | \left(G^{\beta}_{1} \frac{1}{{\cal E}^{\beta}_{0} - h^{\beta}_{0}} \right)^{n_{\beta}}  \left[ \, \, t  \, \, \right.  \label{expenergy} \\
&& \nb \\
&&  \hspace{7cm} \left. \, \, +  \,  \, V \,  \, \right] \left(\frac{1}{{\cal E}^{\alpha}_{0} - h^{\alpha}_{0}} G^{\alpha}_{1} \right)^{n_{\alpha}}  |   \Phi^{\alpha}_{0}  \rangle_{Linked/Irred.}  \, \, \, \, , \nb
\nb
\end{eqnarray}
where the norm is:

\begin{equation}
\langle \, \Theta_{0}| \, \Theta_{0} \rangle   =    \sum_{\alpha,\beta,n_{\alpha},n_{\beta}}  f_{\beta}^{\ast}  f_{\alpha} \, \langle \Phi^{\beta}_{0} | \left(G^{\beta}_{1} \frac{1}{{\cal E}^{\beta}_{0} - h^{\beta}_{0}} \right)^{n_{\beta}} \, \left(\frac{1}{{\cal E}^{\alpha}_{0} - h^{\alpha}_{0}} G^{\alpha}_{1} \right)^{n_{\alpha}}  |   \Phi^{\alpha}_{0}  \rangle_{Linked/Irred.} \, \, \, \, . 
\label{expoverlap}
\end{equation}

The Goldstone diagrams corresponding to Eq.~\ref{expenergy} and Eq.~\ref{expoverlap} are irreducible in terms of the $G^{\beta}$ and $G^{\alpha}$ matrices. The dependence upon the energy parameters $W_{\beta}$ and $W_{\alpha}$ is omitted for convenience. In opposition to the energy expansion in the standard Goldstone-Brueckner theory, Eq.~\ref{expenergy} will provide explicit correlation diagrams associated to the kinetic energy operator.

In the above equations, the diagrammatic expansion is written as a sum over $(\alpha,\beta)$ of several types of vacuum/vacuum diagrams $\langle \Phi^{\beta}_{0} | \, \ldots \,  |   \Phi^{\alpha}_{0}  \rangle$ involving different starting and ending vacua. The sum runs over $k^{2}$ of these terms if $k$ different vacua $| \Phi^{\alpha}_{0}  \rangle$ are considered. Considering a particular couple $(\alpha, \beta)$, let us give rules to calculate the corresponding matrix elements but also to define and draw the new associated diagrams. 

First, all operators having a subscript $\beta$ in Eq.~\ref{expenergy} and~\ref{expoverlap} are written in the corresponding $\left\{\beta\right\}$ single-particle basis which is associated with dashed lines in the diagrams. All operators  having a subscript $\alpha$ are written in the corresponding $\left\{\alpha\right\}$ single-particle basis which is associated with solid lines in the diagrams. The Hamiltonian in the middle of Eq.~\ref{expenergy} is chosen for symmetry reason to be written in the left $\left\{\beta\right\}$ and right $\left\{\alpha\right\}$ single-particle basis:

\begin{equation}
H \, = \, \sum_{m,n} \, t_{\beta_{m} \alpha_{n}} \, \beta^{\dagger}_m \, \alpha_n \, + \, \frac{1}{4} \, \sum_{m,n,p,q} \, \bar{V}_{\beta_{m} \beta_{p} \alpha_{n} \alpha_{q}} \, \beta^{\dagger}_m \, \beta^{\dagger}_p \, \alpha_q \, \alpha_n \, \, \, \, \, , 
\label{Hbasemixte}
\end{equation}
where $\bar{V}_{\beta_{m} \beta_{p} \alpha_{n} \alpha_{q}}$ are antisymmetrized matrix elements.

\begin{figure}
\begin{center}
\leavevmode
\centerline{\psfig{figure=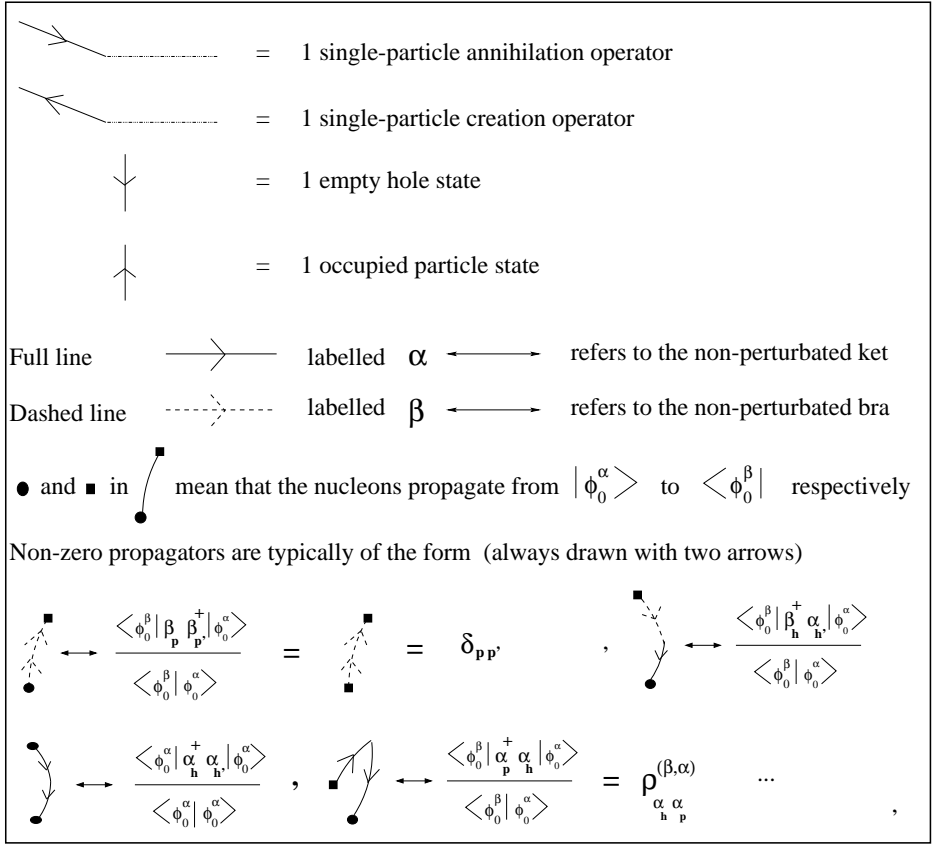,height=13cm}}
\end{center}
\caption{Basic definitions for drawing graphs between different vacua appearing in the generalized energy expansion~\ref{expenergy}. Examples of mixed propagators involved in these graphs are given.}
\label{defgraph}
\end{figure}

Second, one has to insert complete sets of particle-hole excitations between these operators while keeping linked diagrams only. These are particle-hole excitations with respect to $| \Phi^{\alpha}_{0}  \rangle$ around operators with an $\alpha$ subscript, and with respect to $| \Phi^{\beta}_{0}  \rangle$ around operators with a $\beta$ subscript. In order to describe associated graphs, several types of propagators have to be defined depending on their localization in the diagrams. Some of them together with basic definitions are given on Fig~\ref{defgraph}. For example, reading the formula~\ref{expenergy} from right to left, which is equivalent to read diagrams from bottom to top, one can be convinced that the nucleons will first propagate from $|  \Phi^{\alpha}_{0}  \rangle$ to $\langle  \Phi^{\alpha}_{0} |$ in $\{\alpha\}$ individual states. Propagators involving different vacua will appear when going through $H$ because of the complete set of states referring to $|   \Phi^{\alpha}_{0}  \rangle$ inserted on its right and the one referring to $| \Phi^{\beta}_{0}  \rangle$ inserted on its left. After that, they will propagate from $|  \Phi^{\beta}_{0}  \rangle$ to $\langle   \Phi^{\beta}_{0} |$ in $\{\beta\}$ individual states. Graphically, we choose to specify the vacua entering the contraction in a propagator by a square for $|   \Phi^{\beta}_{0}  \rangle$ and by a circle for $|   \Phi^{\alpha}_{0}  \rangle$ at the extremities of the line (see Fig.~\ref{defgraph}). Finally, apart from the presence of two different single-particle basis in the graphs, the use of the Generalized Wick Theorem~\cite{bal1} instead of the standard one~\cite{wick} and thus the apparition of new propagators, other rules for the calculation of diagrams are identical to those defined in Ref.~\cite{gold}. Because of the non-orthogonality between particle-hole excitations built on two different vacua $| \Phi^{\alpha}_{0}  \rangle$ and $| \Phi^{\beta}_{0}  \rangle$, non-zero diagrams are more numerous than usual.

\subsection{Generalized Brueckner Ladders.}
\label{reactionmatrix}

We have now to treat divergences brought about by the $V$ interaction coming from the energy operator $H$ in Eq.~\ref{expenergy}. Their treatment has been separated from the one of the divergences appearing in the time evolution of the wave-function, which have been taken care of through the Brueckner matrices $G^{\alpha}$, because it is of different origin. 

Each diagram of Eq.~\ref{expenergy} involving a $V$ interaction can be associated with diagrams differing only by the fact that a $G^{\alpha}$ interaction connects the original two nucleons state to $V$ through two $\left\{\alpha\right\}$ particle states and/or that a $G^{\beta}$ interaction after $V$ diffuses the two nucleons from $\left\{\beta\right\}$ particle states into the two nucleons final state. This allows to replace $V$ by a $G^{\left(\beta,\alpha\right)}$ matrix summing {\it generalized} particle-particle ladders where nucleons propagate in two types of particle states referring to two different non-orthogonal vacua. In this way, all diagrams originating from Eq.~\ref{expenergy} are generated and each graph given in terms of $G^{\left(\beta,\alpha\right)}$ is well-behaved as will be shown in the next section. The diagram content of the new reaction matrix is illustrated on Fig.~\ref{defgab}.

Analytically,  $G^{\left(\beta,\alpha\right)}$ is expressed in terms of $G^{\alpha}$ and $G^{\beta}$ through:

\begin{eqnarray}
G^{\left(\beta,\alpha\right)} (W_{\beta},W_{\alpha}) \, \, &=&  \, \,  V \, + \, G^{\beta} (W_{\beta}) \, \frac{Q^{\beta}}{W_{\beta} - h^{\beta}_{0}} \, V \nb \\
&& \nb \\
&& \hspace{0.6cm} + \, V \, \frac{Q^{\alpha}}{W_{\alpha} - h^{\alpha}_{0}} \, G^{\alpha} (W_{\alpha}) \,  \nb \\
&& \label{supergmatrix} \\
&& \hspace{0.6cm} + \, \, \, G^{\beta} (W_{\beta}) \, \, \frac{Q^{\beta}}{W_{\beta} - h^{\beta}_{0}} \, \, V \, \, \frac{Q^{\alpha}}{W_{\alpha} - h^{\alpha}_{0}} \, \, G^{\alpha} (W_{\alpha}) \nb \\
&& \nb \\
&=& \, \,  G^{\beta} (W_{\beta}) \, V^{-1} \, G^{\alpha} (W_{\alpha})  \, \, \, \, \, \, , \nb
\end{eqnarray}
the last identity being obtained by grouping terms and using the self-consistent equation satisfied by $G^{\alpha} (W_{\alpha})$ and $G^{\beta} (W_{\beta})$.

The $G^{\left(\beta,\alpha\right)}$ matrix elements together with the definitions of $W_{\beta}$ and $W_{\alpha}$ are given in appendix A.2. $G^{\left(\beta,\alpha\right)}$ also satisfies a self-consistent equation which is given in appendix A.2. Then, the energy expansion can be re-written in terms of $G^{\left(\beta,\alpha\right)}$. The Eq.~\ref{expenergy} becomes:

\begin{figure}
\begin{center}
\leavevmode
\centerline{\psfig{figure=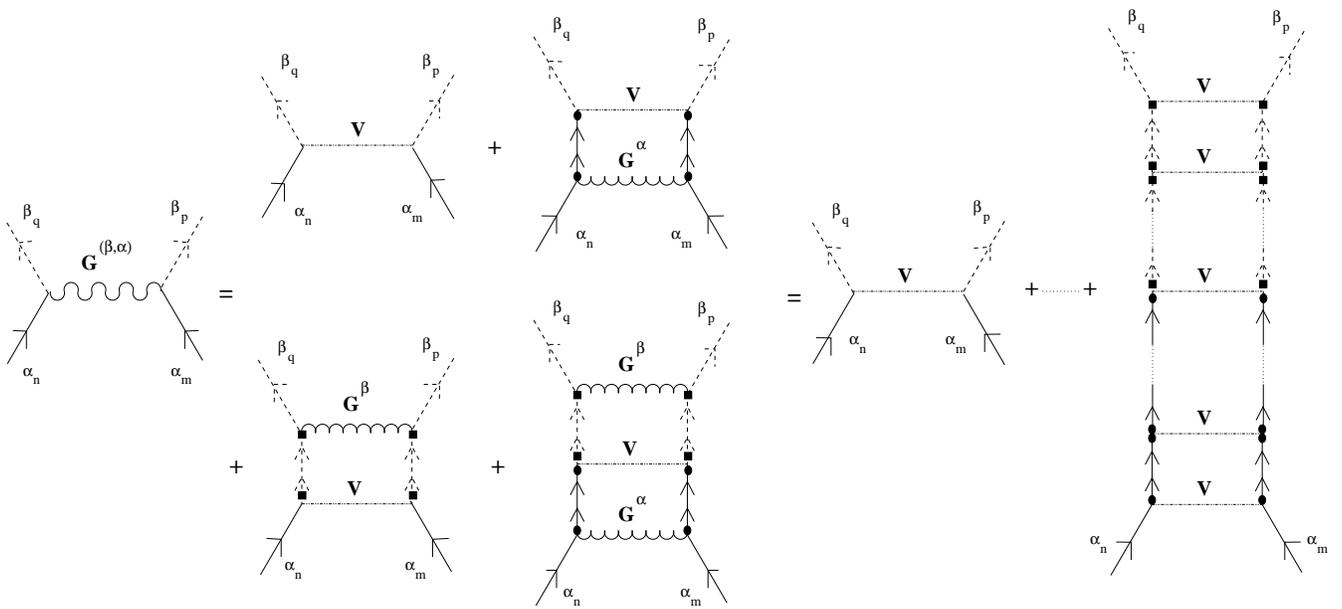,height=8cm}}
\end{center}
\caption{Diagrammatic representation of the $G^{\left(\beta,\alpha\right)}$ matrix summing generalized particle-particle ladders.}
\label{defgab}
\end{figure}

\begin{eqnarray}
&& \langle \, \Theta_{0}| \, H \, | \, \Theta_{0} \rangle \, \, = \,  \sum_{\alpha,\beta,n_{\alpha},n_{\beta}}  f_{\beta}^{\ast}  f_{\alpha} \, \langle \Phi^{\beta}_{0} | \left(G^{\beta}_{1} \frac{1}{{\cal E}^{\beta}_{0} - h^{\beta}_{0}} \right)^{n_{\beta}}  \left[ \, \, t  \, \, \right. \nb \\
&& \label{energywithgab} \\
&& \hspace{6cm} \left. \, \, + \, \,  G^{\left(\beta,\alpha\right)} \, \, \right] \left(\frac{1}{{\cal E}^{\alpha}_{0} - h^{\alpha}_{0}} G^{\alpha}_{1} \right)^{n_{\alpha}}  |   \Phi^{\alpha}_{0}  \rangle_{Linked/Irred.}  \, \, , 
\nb
\end{eqnarray}
where diagrams connecting $G^{\alpha}$ to $G^{\left(\beta,\alpha\right)}$ through two $\left\{\alpha\right\}$ particle states and/or $G^{\left(\beta,\alpha\right)}$ to  $G^{\beta}$ through two $\left\{\beta\right\}$ particle states are excluded in order to avoid double-counting.

The reaction matrix $G^{\left(\beta,\alpha\right)}$ is the effective interaction removing the hard core problem within the framework of our generalized expansion as shown below. A remarkable property of this effective interaction is its dependence with respect to the type of $N$-body matrix element in which it is inserted in Eq.~\ref{energywithgab}. Thus, for $k$ states entering the superposition defined by Eq.~\ref{extended2}, the present scheme requires the calculation of $k$ different $G^{\alpha}$ matrices referring to each vacuum plus $k^2$ $G^{\left(\beta,\alpha\right)}$ matrices referring to each type of $N$-body matrix element. It may be very time-consuming for quantitative calculations. 

As the aim of a Brueckner matrix is to take care of short-range correlations which should not be very sensitive to the single-particle basis used, the matrices $G^{\alpha}$ should be almost identical for all $\alpha$. However, BHF calculations have shown that this is not such a good approximation (see Ref.~\cite{wong} for instance). We will not go further concerning necessary approximations to implement this scheme for quantitative calculations since we stay on a formal level in the present work.

\subsection{Pair Correlated Wave-Function and $G^{\left(\beta,\alpha\right)}$ Matrix Properties.}
\label{2bodycorrfunction}

In standard Brueckner theory, it is useful to illustrate the $G^{\alpha}$ matrix properties through its action on an uncorrelated two-body wave-function~\footnote{Except for the definition of $Q^{\alpha}$, the two-body states are antisymmetrized.}:

\begin{eqnarray}
| \, \alpha_p \, \alpha_q  \rangle \, &=& \, \alpha^{\dagger}_p \, \alpha^{\dagger}_q  \, | \, 0 \,  \rangle  \label{2buncorrfunct} \, \, \, \, \, \, \, .
\end{eqnarray}

This defines a correlated two-body wave-function $| \, \psi^{\alpha}_{pq}  \rangle$~\cite{day} through the equation:

\begin{eqnarray}
V \, | \, \psi^{\alpha}_{pq} (W_{\alpha})  \rangle \, &=& \, G^{\alpha} (W_{\alpha}) \, | \, \alpha_p \, \alpha_q  \rangle \label{2bcorrfunct1} \, \, \, \, \, \, \, ,
\end{eqnarray}
which can be re-written using Eq.~\ref{gmatrix} as:

\begin{eqnarray}
| \, \psi^{\alpha}_{pq} (W_{\alpha})  \rangle \, &=& \, | \, \alpha_p \, \alpha_q  \rangle \, + \,  \frac{Q^{\alpha}}{W_{\alpha} \, \, - \, \, h^{\alpha}_{0}} \, V \, | \, \psi^{\alpha}_{pq} (W_{\alpha})  \rangle \label{2bcorrfunct2} \, \, \, \, \, \, \, .
\end{eqnarray}

It can be shown~\cite{ring1} that the content of the Brueckner summation is such that the correlated pair wave-function  $\psi^{\alpha}_{pq} (W_{\alpha}, \vec{r}, \vec{R})$ vanishes rapidly as the separation distance $\vec{r} \,$ between the two nucleons decreases within the range of the repulsive core. The exclusion of the nucleons from the region of the core in the correlated wave-function makes finite two-body matrix elements of the form:

\begin{eqnarray}
\langle \, \alpha_r \, \alpha_s \, | \,  G^{\alpha} (W_{\alpha}) \, | \, \alpha_p \, \alpha_q  \rangle \, &=& \, \langle \, \alpha_r \, \alpha_s \, | \, V \, | \, \psi^{\alpha}_{pq} (W_{\alpha})  \rangle \label{2bmatrix1} \, \, \, \, \, \, \, .
\end{eqnarray}

This is the physical reason why the standard expansion formulae~\ref{extended} and~\ref{goldenergy} in terms of $G^{\alpha}$ provides well-behaved diagrams. Let us now study the properties of the generalized $G^{\left(\beta,\alpha\right)}$ matrix entering the new energy expansion. As we have symmetrized the role of the bra and the ket in the expansion, the quantity to look at is no longer some correlated two-body ket alone but the two-body matrix element $\langle \, \beta_r \, \beta_s | \, G^{\left(\beta,\alpha\right)} \, | \, \alpha_p \, \alpha_q  \rangle$ itself.

Using Eq.~\ref{supergmatrix} together with the definition~\ref{2bcorrfunct1} for the pair correlated wave-function, one can write

\begin{eqnarray}
\langle \, \beta_r \, \beta_s \, | \,  G^{\left(\beta,\alpha\right)} (W_{\alpha},W_{\beta}) \, | \, \alpha_p \, \alpha_q  \rangle \, &=& \,  \langle \, \beta_r \, \beta_s \, | \, G^{\beta} (W_{\beta}) \, V^{-1} \, G^{\alpha} (W_{\alpha})  \, | \, \alpha_p \, \alpha_q  \rangle \nb    \\
&& \label{2bmatrix3} \\
&=& \, \langle \, \psi^{\beta}_{rs} (W_{\beta}) | \, V \,| \, \psi^{\alpha}_{pq} (W_{\alpha})  \rangle \, \, \, \, .
\nb
\end{eqnarray}

This identity shows that $G^{\left(\beta,\alpha\right)}$ is well defined to renormalize two-body short-range correlations stemming from the repulsive core of the interaction. It also shows how the energy expansion~\ref{goldenergy} provides well-behaved diagrams. Comparing Eq.~\ref{2bmatrix3} with Eq.~\ref{2bmatrix1}, one sees that $G^{\left(\beta,\alpha\right)}$ renormalizes more correlations than standard Brueckner matrices since it correlates both the two-body ket and the two-body bra. This originates from writting the ground-state energy as the expectation value of the hamiltonian in the actual ground-state and resumming the ladders accordingly, instead of taking the scalar product of $H \, | \, \Theta_{0} \rangle$ with a vacuum and resumming the ladders as done in the standard Brueckner theory.

Given the above rules and renormalized interactions $G^{\alpha}$, $G^{\beta}$ and $G^{\left(\beta,\alpha\right)}$, Eq.~\ref{energywithgab} together with the norm provide a perturbative formula for the ground-state energy. It is worth mentioning that all the exponents $(n_{\beta}, n_{\alpha})$ defining the orders of the expansion for each $N$-body matrix element in Eq~\ref{energywithgab} and Eq.~\ref{expoverlap} are all taken equal $n_{\alpha} = n_{\beta} = n \, \, \, , \,  \forall \, \, (\alpha, \beta)$. This means that, at a given order, identical diagrams with respect to each unperturbed state will be dropped both in the ket and in the bra\footnote{{\it Once the hard-core problem is solved}, the meaningful order in the energy expansion~\ref{energywithgab} is defined as the number of explicit $G^{\alpha}$ interactions associated with each term originating from the ket (or the bra).}. The order at which these expansions are truncated must be the same for $\langle \, \Theta_{0} | H | \, \Theta_{0}  \rangle$ and $\langle \, \Theta_{0} | \, \Theta_{0}  \rangle$.

\subsection{Lowest-Order Approximation.}
\label{zeroorder}

\begin{figure}
\begin{center}
\leavevmode
\centerline{\psfig{figure=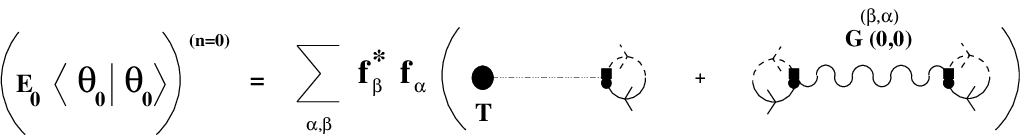,height=1.7cm}}
\end{center}
\caption{Lowest-order approximation for the energy in terms of the $G^{\left(\beta,\alpha\right)}$ matrix. Square and dot symbols are defined in Fig.~\ref{defgraph}.}
\label{finalenergy}
\end{figure}

Let us consider the lowest-order approximation. It corresponds to retain the $n = 0$ term only in Eq.~\ref{expoverlap} and Eq.~\ref{energywithgab}. The approximate energy of the interacting system becomes:

\begin{equation}
E^{\, n=0}_{0} \, = \, \frac{\sum_{\alpha,\beta} \, f_{\beta}^{\ast}  f_{\alpha}   \langle \, \Phi^{\beta}_{0} | \, t \, + \, G^{\left(\beta,\alpha\right)} (0,0) \, | \,  \Phi^{\alpha}_{0}  \rangle}{\sum_{\alpha,\beta}^{} \, f_{\beta}^{\ast}  f_{\alpha}  \langle \, \Phi^{\beta}_{0} |  \,  \Phi^{\alpha}_{0}  \rangle} \, \, \, \, \, ,
\label{lastenergy}
\end{equation}
which is identical to Eq.~\ref{energy1} except that the starting two-body interaction $V$ has been replaced by the $G^{\left(\beta,\alpha\right)}$ matrix on the energy shell ($W_{\alpha} = W_{\beta} = 0$). Consequently, we have achieved the goal to provide general configurations mixing calculations with a well-defined perturbative equivalent in terms of an effective interaction removing the hard-core problem. The corresponding diagrammatic picture is shown on Fig.~\ref{finalenergy}. This lowest-order already contains long-range correlations through the mixing of non-orthogonal vacua. This suggests that the presently developed perturbation theory should be appropriate for systems being soft with respect to some collective degree of freedom since the mixing of several mean-fields is an efficient way to take the associated correlations into account. Within this framework, the lowest-order corresponds to an adiabatic Born-Oppenheimer approximation justified when the individual degrees of freedom relax much faster than the collective one. Going to higher orders in the expansion allows to include diabatic effects in the coupling of these two kinds of degrees of freedom.

A mean-field approximation is recovered by taking a single $f_{\alpha_0}$ coefficient to be different from zero. In this case, Eq.~\ref{lastenergy} contains a diagonal matrix element only and $G^{\left(\beta,\alpha\right)}$ reduces to a single term $G^{\left(\alpha_0,\alpha_0\right)}$. In comparison with Eq.~\ref{middleenergy}, the effective interaction appearing at the mean-field level in the present case is different from the standard one. As shown in section~\ref{2bodycorrfunction}, $G^{\left(\alpha_0,\alpha_0\right)}$ generates two-body matrix elements including more correlations than $G^{\alpha_0}$ does. In some way, our scheme is not a simple extension of the standard Goldstone formulae since we do not recover it for $f_{\alpha} = \delta_{\alpha \alpha_0} \, f_{\alpha_0}$. Looking at this particular case, one could think that the expression of the energy~\ref{expenergy} together with the $G^{\left(\alpha_0,\alpha_0\right)}$ matrix lead to some double-counting with respect to the standard scheme. The Eq.~\ref{expenergy} being exact, this is simply a different way of resumming two-body correlations into the interaction when expanding the whole energy instead of some correlation energy. It may appear as an unnecessary complication when considering a single Fermi-sea but is the simplest way of superposing several of them to get the energy as given by Eq.~\ref{lastenergy} at the lowest-order in $G^{\beta}$ and $G^{\alpha}$. This was our main goal.

The matrix $G^{\left(\beta,\alpha\right)}$ depends upon the matrix element $\langle \, \Phi^{\beta}_{0} | \, \ldots \, | \, \Phi^{\alpha}_{0}  \rangle$ in which it is inserted. Thus, the energy at the lowest-order cannot be factorize into a form as given by Eq.~\ref{factorisee} which is built from a defined state $| \, \Psi_{k}  \rangle$ (see Eq.~\ref{mixing}). Within the perturbative calculation renormalizing short-range correlations, $| \, \Psi_{k}  \rangle$ should not be considered as the corresponding approximate wave-function of the actual ground-state of the system from which other observables could be calculated together with bare operators. Actually, this is only by analogy with a true variational calculation making use of the starting Hamiltonian that one would use $| \, \Psi_{k}  \rangle$ as the corresponding approximate state of the system. The calculation done in this way for approximate values of other observables will be of the same quality as for the energy at a given order only if short-range correlations associated with the corresponding operator are small. This is precisely the situation which has been encountered in the GCM and the projected mean-field method when using phenomenological interactions depending on the mixed local density $\rho^{\left(\beta,\alpha\right)} (\vec{R})$~\cite{bonche2,bonche3,haider,heenen}. Indeed, while this interaction depends on the matrix element $\langle \, \Phi^{\beta}_{0} | \, \ldots \, | \, \Phi^{\alpha}_{0}  \rangle$, the state $| \, \Psi_{k}  \rangle$ has been used to calculate observables such as probability transitions. We will thus consider this approximation.

The reason why no perturbatively well defined approximate state can be extracted from Eq.~\ref{lastenergy} is related to the fact that some diagrams are summed in $G^{\left(\beta,\alpha\right)}$ incoherently with respect to a well-defined order in the expansion of the state $| \, \Theta_{0}  \rangle$. We have actually defined a perturbative expansion for the mean value of an observable (energy) without taking care of the state of the system itself. It is worth noting that defining an approximate energy through the mean value of $H = t + V$ in any approximate form of $| \, \Theta_{0}  \rangle$ would lead to divergent diagrams. Indeed, $V$ could not systematically enter a ladder in this case. That is why ladder resummations has to be performed in the energy before any approximation takes place\footnote{The situation is similar when dealing with approximate forms of $| \, \Theta_{0}  \rangle$ in $E_{0} = {\cal E}_{0}^{\alpha} + \Delta {\cal E}_{0}^{\alpha} = \langle \, \Phi^{\alpha}_{0} | \, H \, | \, \Theta_{0}  \rangle$ in the standard Brueckner-Goldstone theory. Then, $| \, \Phi^{\alpha}_{0}  \rangle$  is not rigourously defined to be the approximate state associated with the BHF energy.}.

The same kind of situation is encountered in usual shell-model calculations making use of effective operators in a limited valence space. In fact, this expresses the fact that the nuclear mean-field or beyond mean-field energy in calculations making use of phenomenological effective forces is more to be seen as a functional (of local densities in general) than as the mean value of a two-body Hamiltonian in a defined state. We anticipate the extension of the Skyrme functional beyond the vicinity of a single mean-field~\cite{hagino} which will be proposed in a forthcoming publication. This extension relying on Eq.~\ref{lastenergy}, a formal link is kept with the bare interaction. Contrary to Kohn-Sham philosophy~\cite{kohn}, this strategy is by nature systematic and does not aim at incorporating all the missing correlations at the considered order.

The scheme developed up to now is valid whatever the coefficients $f_{\alpha}$ are. While they are of no importance as long as no approximation is done, they will influence the results at all orders of the expansion. Let us now study two applications of the general scheme associated with two different choices of these coefficients.

\section{GCM.}
\label{gcm}

Once the set of product states is chosen along a collective deformation path, the perturbative energy can be minimized with respect to $f^{\ast}_{\beta}$ coefficients at any order of the expansion. At the lowest order, the minimization of $E^{\,n=0}_{0} [f^{\ast}_{\beta}, f_{\alpha}]$ is equivalent to the eigenvalue problem:

\begin{equation}
\left(\begin{array}{c}
		\\
	 \langle \, \Phi^{\beta}_{0} |  t +  G^{\left(\beta,\alpha\right)}  | \,  \Phi^{\alpha}_{0}  \rangle \\
		\\
	\end{array}\right) 
\left(\begin{array}{c}
		\\
	f_{\alpha} \\
		\\
	\end{array}\right) = E^{\,n=0}_{0} 
\left(\begin{array}{c}
		\\
	 \langle \, \Phi^{\beta}_{0} | \, \Phi^{\alpha}_{0}  \rangle \\
		\\
	\end{array}\right) 
\left(\begin{array}{c}
		\\
	f_{\alpha} \\
		\\
	\end{array}\right) \, \, \, \, \, \, \, , \label{tableaugcm}
\end{equation}
written in matrix form because of the dependence of the effective ``Hamiltonian'' $t \, + \, G^{\left(\beta,\alpha\right)} (0,0)$ on the matrix element $\langle \, \Phi^{\beta}_{0} | \, \ldots \, | \, \Phi^{\alpha}_{0}  \rangle$. 

These are the Hill-Wheeler equations encountered in the GCM~\cite{hill, ring1} written in the non-orthogonal basis $\{| \, \Phi^{\alpha}_{0}  \rangle\}$ in terms of the appropriate effective interaction. This result motivates the GCM from the perturbative point of view for the first time. Let us emphasize that we are in no way referring to the Ritz variational principle here since, as already mentioned in the previous section, the Eq.~\ref{lastenergy} is not obtained through the expectation value of the actual Hamiltonian into an well defined approximate state. In particular, the corresponding energy can be lower than the actual ground-state energy. Finally, minimizing the energy at some higher orders in the expansion motivates the introduction of diabatic effects in the GCM~\cite{diet,tajima1,tajima2}.

Within the present interpretation, the diagonalization of the GCM matrix provides several sets of ``eigenvectors'' $\left\{f_{\alpha}^{k}\right\}$, $k=0,1,2,..$ corresponding to different approximations of the ground-state energy~\ref{lastenergy}. Everything having been written from the outset for the ground-state, no excited state is obtained here. The corresponding $| \, \Psi_{k}  \rangle$ form a set of orthogonal states where each of them is not orthogonal to the actual ground-state $| \, \Theta_{0} \rangle$ which it approximates. Using a simple argument, it is however possible to interpret the $| \, \Psi_{k}  \rangle$, $k \geq 1$, as good approximations of the excited states of the system. Indeed, while $| \, \Psi_{0}  \rangle$ has a maximum overlap with $| \, \Theta_{0} \rangle$, $| \, \Psi_{k}  \rangle$ which is orthogonal to $| \, \Psi_{0}  \rangle$ will have a large overlap with the eigenstate $| \, \Theta_{k} \rangle$ corresponding to the same ordering of the energies.

\section{Symmetry Restoration.}
\label{proj}

The second case of interest consists in concentrating on the approximate state rather than on the approximate energy of the system. The actual ground-state $| \, \Theta^{\alpha}_{0}  \rangle$ defined through Eq.~\ref{goldstone} obeys the symmetries of the system whatever those broken by the product state $| \, \Phi^{\alpha}_{0} \rangle$ from which it develops at $t = -\infty$. However, a problem arises when diagrams are dropped in the expansion, since then, symmetry breaking may appear in the wave-function. For instance, if the $n \, = \, 0$ term only is retained, one recovers the symmetry breaking mean-field solution $| \, \Phi^{\alpha}_{0} \rangle$. Another way to stress this point is to remember that usual perturbative corrections with respect to $| \, \Phi^{\alpha}_{0} \rangle$ will not restore any broken symmetry unless an infinite number of diagrams is summed\footnote{In the particular case of rotational invariance, one could start from a spherical mean-field and use Bloch-Horowitz perturbation theory for degenerate unperturbed states~\cite{bloch2} if dealing with an open-shell nucleus. As we are interested in relating the projected mean-field method to a perturbative expansion, we omit this possibility and continue with deformed mean-field unperturbed states.}~\cite{blaiz.ripk}.

In variational calculations, it is well-known that such an infinity can be taken care of through a linear superposition of symmetry breaking states $| \, \Phi^{\alpha}_{0} \rangle$ as given by Eq.~\ref{mixing}. This is the projected mean-field method. According to the symmetry, it corresponds to a particular choice of the unperturbed states $| \, \Phi^{\alpha}_{0} \rangle$ and mixing coefficients $f^{\alpha}$. Let us now show how our extended perturbation theory motivates such a method by restoring good quantum numbers in the standard Goldstone-Brueckner expansion. We examplify this through the restoration of rotational invariance.

We define $| \, \Phi^{0}_{0} \rangle$ as the non-degenerate ground-state of an axially symmetric deformed one-body Hamiltonian $h^{0}_{0}$. A set of rotated states around the y axis orthogonal to the symmetry axis is obtained through~:

\begin{equation}
| \, \Phi^{\alpha}_{0} \rangle \, =  \, \exp \left(i \, \frac{\pi \alpha}{n} \, J_y  \right)  \, | \, \Phi_{0}^{0} \rangle \, = \, R (\alpha) \, | \, \Phi_{0}^{0} \rangle  \, \, \, \, , 
\label{rot}
\end{equation}
with $\alpha$ varying from $-n$ to $n$ as an integer. The Hamiltonian being invariant under this tranformation, the rotating operator $R (\alpha)$ commutes with it. $| \, \Phi^{0}_{0} \rangle$ being $h^{0}_{0}$'s non-degenerate ground-state with the energy ${\cal E}^{0}_{0}$, $| \, \Phi^{\alpha}_{0} \rangle$ will be the one of the rotating Hamiltonian $h^{\alpha}_{0} = R (\alpha) \, h^{0}_{0} \,R^{\dagger} (\alpha)$ with the same eigenenergy. Thus, the adiabatic evolution operator is written in the $\alpha$ interaction representation under the form:

\begin{eqnarray}
U^{\alpha}_{\epsilon} (t,t_0) \, &=& \, e^{i h^{\alpha}_0 t / \hbar} \, \, U_{\epsilon} (t,t_0) \, \,  e^{- i h^{\alpha}_0 t_0 / \hbar}  \nb \\
&& \nb \\
&=& \, R (\alpha) \, \, e^{i h^{0}_0 t / \hbar} \, \, R^{\dagger} (\alpha) \, \, e^{- i \int_{t_0}^{t} H^{\alpha} (\epsilon,\tau) d\tau  / \hbar} \, \, R (\alpha) \, \, e^{- i h^{0}_0 t_0 / \hbar} \, \, R^{\dagger} (\alpha)  \nb \\
&& \nb \\
&=& \,  R (\alpha) \, \, U^{0}_{\epsilon} (t,t_0) \, \, R^{\dagger} (\alpha)  \nb 
\label{Uevolproj} 
\end{eqnarray}
because the $\alpha$ dependence of the auxiliary Hamiltonian is:

\begin{equation}
H^{\alpha} (\epsilon,\tau) \, = \, R (\alpha) \, \, H^{0}(\epsilon,\tau) \, \, R^{\dagger} (\alpha) \, \, \, \, \, \, \, \, .
\label{enieme}
\end{equation}

Consequently, the ground-state~\ref{extended2} can be re-written as:

\begin{eqnarray}
| \, \Theta_{0} \rangle \, &=& \lim_{\epsilon \, \rightarrow \, 0} \sum_{\alpha} \, f_{\alpha} \, \frac{U^{\alpha}_{\epsilon} (0,-\infty) \, | \, \Phi^{\alpha}_{0} \rangle}{\langle \, \Phi^{\alpha}_{0}\, | U^{\alpha}_{\epsilon} (0,-\infty) | \, \Phi^{\alpha}_{0} \rangle } \nb \\
&& \nb \\
&=& \,\left[ \sum_{\alpha}^{} \, f_{\alpha} \, R (\alpha) \right] \, \left[ \lim_{\epsilon \, \rightarrow \, 0} \, \frac{U^{0}_{\epsilon} (0,-\infty) \, | \, \Phi^{0}_{0} \rangle}{\langle \, \Phi^{0}_{0}\, | U^{0}_{\epsilon} (0,-\infty) | \, \Phi^{0}_{0} \rangle } \right] \label{extended3} \\
&&  \nb \\
&=& \, \left[ \sum_{\alpha}^{} \, f_{\alpha} \, R (\alpha) \right] \, \left[ \sum_{n}^{} \, \left(\frac{1}{{\cal E}^{0}_{0} \, - \, h^{0}_{0}} \, G^{0}_{1} \right)^n \, | \, \Phi^{0}_{0} \rangle_{Linked/Irred.} \right] \, \, \, \, \, \, \, \, . \nb
\end{eqnarray}

If $f_{\alpha}$ satisfies~\cite{ring1}:

\begin{equation}
f^{I0}_{\alpha} \,  = \, \frac{2I + 1}{2n} \, sin(\pi \alpha / n) \, d^{I \, \ast}_{00} (\pi \alpha / n) \, \, \, \, \, , \label{coeffprojJ}
\end{equation}
where $d^{I}_{M0}$ is the Wigner function for $(I,M,K=0)$ quantum numbers~\cite{blaiz.ripk}, the operator:

\begin{equation}
\hat{P}_{I0} \, = \, \sum_{\alpha= -n}^{n} \, f^{I0}_{\alpha} \, R (\alpha) \, \, \, \, \, ,
\label{operatorproj}
\end{equation}
is the projector on total spin $I$ and projection on the $z$ axis $M=0$\footnote{The projector for $M\neq0$ or for a triaxially deformed product state $| \, \Phi^{0}_{0} \rangle$ is more complicated but the calculation can be extended without major difficulty.}. Thus, the Eq.~\ref{extended3} becomes:

\begin{equation}
| \, \Theta_{0} \rangle \, = \, \hat{P}_{I0} \, | \, \Theta^{0}_{0}  \rangle \, \, \, \, \, ,
\label{projground-state}
\end{equation}
the projection being trivial if no approximation is done on $| \, \Theta^{0}_{0}  \rangle$. For an even-even nucleus, the projection gives 0 for $I \neq 0$ and the identity for $I = 0$.

Finally, the extended perturbation theory allows to restore broken symmetries in the standard Goldstone-Brueckner expansion by projecting the approximate state. This formulation motivates the standard projected mean-field method from a perturbative point of view for the first time and generalizes it to any order of the expansion. Moreover, the method allows for a coherent summation of Brueckner ladders through the use of rotated Brueckner matrices $R (\alpha) \, G^{0} \, R^{\dagger} (\alpha)$ for each component of the approximate wave-function.

Let us now study the corresponding expansion of the energy. As $\hat{P}_{I0}$ commutes with $H$ and satisfies $\hat{P}^{2}_{I0} \, = \, \hat{P}_{I0}$, the actual ground-state energy defined by Eq.~\ref{expenergy} becomes:

\begin{equation}
E_{0}  \, = \, \frac{\langle \, \Theta^{0}_{0} | \, H \, \hat{P}_{00} \, | \, \Theta^{0}_{0}  \rangle}{\langle \, \Theta^{0}_{0} | \, \hat{P}_{00} \, | \, \Theta^{0}_{0}  \rangle} \, \, \, \, \, ,
 \label{energyprojection} 
\end{equation}
which includes a simple sum over $\alpha$. In order to get the form~\ref{energywithgab}, the divergences associated with the $V$ interaction coming from $H$ have to be regularized. The generalized Brueckner matrix doing so reads as:

\begin{eqnarray}
G^{\left(0,\alpha\right)} (W_{0}',W_{0}) \, \, &=&  \, \,  V \, + \, G^{0} (W_{0}') \, \frac{Q^{0}}{W_{0}' - h^{0}_{0}} \, V  \nb \\
&& \, \nb \\
&& \hspace{0.6cm} + \, V \, R (\alpha) \, \frac{Q^{0}}{W_{0} - h^{0}_{0}} \, G^{0} (W_{0}) \, R^{\dagger} (\alpha) \nb \\
&& \label{supergmatrixproj}\\
&& \hspace{0.6cm} +  \, G^{0} (W_{0}') \, \, \frac{Q^{0}}{W_{0}' - h^{0}_{0}} \, \, V \, R (\alpha) \, \, \frac{Q^{0}}{W_{0} - h^{0}_{0}} \, \, G^{0} (W_{0}) \, R^{\dagger} (\alpha) \, \, \, \, \, \, \, ,
\nb
\end{eqnarray}
and depends on the matrix element $\langle \, \Theta^{0}_{0} | \, \ldots \, R (\alpha) \, | \, \Theta^{0}_{0}  \rangle$ in which it is inserted. Thus, the projection presents a simplification with respect to the general case since a single standard $G^{\alpha = 0}$ matrix and $k$ $G^{\left(0,\alpha\right)}$ matrices have to be calculated explicitely. Once two-body correlations are summed in the interaction, the energy  at the lowest order in $G^{\alpha = 0}$ is written:

\begin{equation}
E^{\, n = 0}_{0}  \, = \, \frac{\sum_{\alpha= -n}^{n} \, f^{00}_{\alpha} \, \langle \, \Phi_{0}^{0} | \, \left[ t \, + \, G^{\left(0,\alpha\right)} (0,0)\right] \, R (\alpha) \, | \, \Phi_{0}^{0}  \rangle}{\sum_{\alpha= -n}^{n} \, f^{00}_{\alpha} \,  \langle \, \Phi_{0}^{0} | \, R (\alpha) \, | \,\Phi_{0}^{0}   \rangle} \, \, \, \, \, .
\label{projection0order}
\end{equation}

It takes the form of the projected mean-field energy on spin $I = 0$ with the appropriate effective interaction $G^{\left(0,\alpha\right)} (0,0)$. Two remarks still have to be done.

In agreement with the discussion given in section~\ref{zeroorder}, the correlated wave-function~:

\begin{equation}
| \, \Psi_{0}  \rangle \, = \, \hat{P}_{00} \, | \,\Phi_{0}^{0} \rangle \, \, \, \, \, \, ,
\label{projstate0order}
\end{equation}
is taken as the approximate ground-state of the system corresponding to the energy~\ref{projstate0order}.

Even if the projection of $| \, \Theta^{0}_{0}  \rangle$ for $I \neq 0$ is zero, this is not necessarily true for an approximation of this state. For the same reasons of orthogonality as in the GCM, it is natural to use Eq.~\ref{projection0order} and~\ref{projstate0order} for $I \neq 0$ as good approximations of the energy and the wave-function of the excitated states $| \, \Theta_{I} \rangle$ of the system.

\section{Conclusions.}
\label{conclu}

We have derived an extended Brueckner-Goldstone perturbative expansion for the ground-state energy of an interacting system. This scheme allows to recover at the lowest-order the energy of a mixing of non-orthogonal product states in variational calculation. The expansion is valid for systems interacting through a strongly repulsive interaction at short distances as it is written in terms of a generalized Brueckner matrix renormalizing two-body short-range correlations in the appropriate way.

The main achievement of the present work is thus to provide a link between the generator coordinate or the projected mean-field methods making use of (phenomenological) effective interactions and a perturbative expansion of the actual ground-state energy of the system.  In particular, our method allows to understand the usual projected mean-field method as the lowest-order approximation of a symmetry restored perturbation theory.

In addition, this work formally provides a well-defined effective interaction removing the hard core problem to be used the GCM or the projected mean-field method. In particular, the summation of particle-particle ladders is adapted to the superposition of non-orthogonal vacua characterizing these two methods. The link between the presently defined effective interaction and phenomenological interactions used in configuration mixing calculations of finite nuclei such as the Gogny or the Skyrme forces is the aim of a forthcoming publication. A new prescription for their density-dependence beyond the mean-field relying on the presently developed perturbation theory will then be tested on quantitative configuration mixing calculations in finite nuclei. Let us say that the feasibility of perturbative calculations using this extended Brueckner-Goldstone theory to deal with large amplitude collective motions may be questionable at the present time. This is a reason why we have planned in a near future to use it mainly as a tool to get theoretically grounded guidelines for density-dependent phenomenological interactions to be used in beyond mean-field calculations.

\section{Acknowledgments.}
\label{secremer}

The author is very greatful to G. Ripka for fruitful discussions during the elaboration of this work and to P. Bonche and P.-H. Heenen for a careful reading of the manuscript.

\begin{appendix} 

\section{$G^{\alpha}$ Matrix Elements and Starting Energy.}
\label{galpha}

An explicit matrix element of $G^{\alpha}$ as defined by Eq.~\ref{gmatrix} is:

\begin{eqnarray}
G^{\alpha}_{\alpha_m  \alpha_n \alpha_p \alpha_q}  (W_{\alpha}) \, &=& \, \hspace{3cm}  V_{\alpha_m  \alpha_n \alpha_p \alpha_q} \nb \\
&& \,   \label{defomega1} \\
&& + \, \, \frac{1}{2} \, \sum_{\epsilon_{\alpha_{r}}, \epsilon_{\alpha_{s}}  > \epsilon^{\alpha}_{F}}^{} \, \frac{V_{\alpha_m \alpha_n \alpha_r \alpha_s} \, \, \, G^{\alpha}_{\alpha_r  \alpha_s \alpha_p \alpha_q} (W_{\alpha})}{\epsilon_{\alpha_p}\!+\!\epsilon_{\alpha_q}\!-\!\epsilon_{\alpha_r}\!-\!\epsilon_{\alpha_s}\!-\!W_{\alpha}}  \, \, \, \, \, \, \, \, \, \, \, \, \, \, \, .
\nb
\end{eqnarray}

In this definition, $W_{\alpha}$ corresponds to the total excitation energy of the intermediate state before the $G^{\alpha}$ interaction. In Eq.~\ref{defomega1}, we took non-antisymme\-tri\-zed matrix elements for $V$ which brings about the necessity to antisymmetrize those of $G^{\alpha}$ at the end of the calculation. The other possibility is to start from the outset with antisymmetrised matrix elements of $V$.

\section{$G^{\left(\beta,\alpha\right)}$ Matrix Elements and Starting Energies.}
\label{gbetaalpha}

The reaction matrix $G^{\left(\beta,\alpha\right)}$ satisfies a self-consistent equation generalizing the one for the usual $G^{\alpha}$ Brueckner matrix:

\begin{eqnarray}
G^{\left(\beta,\alpha\right)} (W_{\beta},W_{\alpha}) \, \, &=&  \, \, V \, \, \, + \, \, \, G^{\left(\beta,\alpha\right)} (W_{\beta},W_{\alpha}) \, \, \frac{Q^{\alpha}}{W_{\alpha} - h^{\alpha}_{0}} \, \, V \, \nb \\
&& \, \label{supergmatrixself}\\
&& \, \, +  \, \, \, \, V \, \, \frac{Q^{\beta}}{W_{\beta} - h^{\beta}_{0}} \, \, G^{\left(\beta,\alpha\right)} (W_{\beta},W_{\alpha}) \, \nb \\
&& \, \nb \\
&& \, \, - \, \, \, \, V \, \, \frac{Q^{\beta}}{W_{\beta} - h^{\beta}_{0}} \, \, G^{\left(\beta,\alpha\right)} (W_{\beta},W_{\alpha}) \, \, \frac{Q^{\alpha}}{W_{\alpha} - h^{\alpha}_{0}} \, \, V  \, \, \, \,  \, \, \, \, \, \, \, \, . \nb 
\end{eqnarray}

Then, an explicit matrix element is

\begin{eqnarray}
G^{\left(\beta,\alpha\right)}_{\beta_m  \beta_n \alpha_p \alpha_q} (W_{\beta},W_{\alpha}) \, &=&  \hspace{2cm} V_{\beta_m  \beta_n \alpha_p \alpha_q}  \label{defomega2} \\
&& \, \nb \\
&& + \, \, \frac{1}{2} \, \sum_{\epsilon_{\alpha_{r}}, \epsilon_{\alpha_{s}}  > \epsilon^{\alpha}_{F}}^{} \, \frac{G^{\left(\beta,\alpha\right)}_{\beta_m  \beta_n \alpha_r \alpha_s}  (W_{\beta},W_{\alpha}) \, \, \, V_{\alpha_r \alpha_s \alpha_p \alpha_q}}{\epsilon_{\alpha_p}\!+\!\epsilon_{\alpha_q}\!-\!\epsilon_{\alpha_r}\!-\!\epsilon_{\alpha_s}\!-\!W_{\alpha}}  \, \nb \\
&& \nb \\
&& + \, \, \frac{1}{2} \, \sum_{\epsilon_{\beta_{t}}, \epsilon_{\beta_{u}}  > \epsilon^{\beta}_{F}}^{} \, \frac{V_{\beta_m  \beta_n \beta_t  \beta_u}  \, \, \, G^{\left(\beta,\alpha\right)}_{\beta_t  \beta_u \alpha_p \alpha_q} (W_{\beta},W_{\alpha})}{\epsilon_{\beta_m}\!+\!\epsilon_{\beta_n}\!-\!\epsilon_{\beta_t}\!-\!\epsilon_{\beta_u}\!-\!W_{\beta}} \nb \\
&& \, \nb \\
&& - \, \, \frac{1}{4} \, \sum_{\mathrel{\mathop{\epsilon_{\beta_{t}}, \epsilon_{\beta_{u}}  > \epsilon^{\beta}_{F}}\limits_{\epsilon_{\alpha_{r}}, \epsilon_{\alpha_{s}}  > \epsilon^{\alpha}_{F}}}}^{} \, \frac{V_{\beta_m  \beta_n \beta_t  \beta_u} \, \, \, G^{\left(\beta,\alpha\right)}_{\beta_t  \beta_u \alpha_r \alpha_s}  (W_{\beta},W_{\alpha}) \, \, \, V_{\alpha_r \alpha_s \alpha_p \alpha_q}}{\left(\epsilon_{\beta_m}\!+\!\epsilon_{\beta_n}\!-\!\epsilon_{\beta_t}\!-\!\epsilon_{\beta_u}\!-\!W_{\beta}\right) \, \left(\epsilon_{\alpha_p}\!+\!\epsilon_{\alpha_q}\!-\!\epsilon_{\alpha_r}\!-\!\epsilon_{\alpha_s}\!-\!W_{\alpha}\right)}\, \, \, \, \, \, \, \, \, \, \, \, \, . \nb
\end{eqnarray}

In this definition, $W_{\alpha}$ corresponds to the total excitation energy of the intermediate state before the $G^{\left(\beta,\alpha\right)}$ interaction while $W_{\beta}$ corresponds to the total excitation energy of the intermediate state after the $G^{\left(\beta,\alpha\right)}$ interaction.

\end{appendix} 

\clearpage


\end{document}